\documentclass[twocolumn,showpacs,preprintnumbers,amsmath,amssymb]{revtex4}
\usepackage{graphicx}
\usepackage{bm}
\usepackage{dcolumn}

\def\prl#1#2#3{{ Phys.   Rev.   Lett.  } {\bf #1}, #2 (#3)}
\def\pla#1#2#3{Phys.   Lett.   A {\bf #1}, #2 (#3)}

\def\pre#1#2#3{Phys.   Rev.   E {\bf #1}, #2 (#3)}
\def\prb#1#2#3{Phys.   Rev.   B {\bf #1}, #2 (#3)}

\def\pra#1#2#3{Phys.   Rev.   A {\bf #1}, #2 (#3)}

\def\jpa#1#2#3{J.   Phys.   A {\bf #1}, #2 (#3)}
\def\jsp#1#2#3{J.   Stat.   Phys.   {\bf #1}, #2 (#3)}

\def\rmp#1#2#3{Rev.   Mod.   Phys.   {\bf #1}, #2 (#3)}

\def\noi{\noindent}
\def\bc{\begin{center}}
\def\ec{\end{center}}
 \newcommand{\bea}{\begin{equation}}
 \newcommand{\eea}{\end{equation}\noi}
 \newcommand{\ber}{\begin{eqnarray}}
 \newcommand{\eer}{\end{eqnarray}\noi}
\begin{document}
\title{ DIFFUSION IN MODULATED  MEDIA}
\author{Himadri    S.   Samanta}\email{tphss@mahendra.iacs.res.in}
\affiliation{Department of Theoretical Physics,
Indian Association for the Cultivation of Science \\
Jadavpur, Calcutta 700 032, India}
\date{\today}
\begin{abstract}
We study the motion of Brownian particle in modulated media in the strong damping limit by using {\em toy model}, with special emphasis 
on the transition from localise to diffusive behavior. By using model potential we have seen the localised behavior when the number of minima of the potential is finite in the asymptotic time limit. In the limit of infinite
number of minima we have seen the diffusive behavior.We calculate exactly the diffusion coefficient in periodic field of force.  We have also studied the transport in commensurate and incommensurate media.

\end{abstract}

\newpage

\maketitle

\section{Introduction}
Thermally activated barrier crossing has been subject of research for many decades since 
the pioneering work of Kramers\cite{a1} on the subject. A fair amount of attention has recently been devoted to the study of complex non-equilibrium systems. These include the case of diffusion 
over a barrier in the presence of harmonic force\cite{a2} and the diffusion over a fluctuating barrier\cite{a3,a3a4}.
The hallmark of the former situation is the phenomenon of stochastic resonance. The problem of surmounting potential barriers\cite{a3a4,a4} has gained importance in other field of science such as evolutionary computations and global optimization as well. In recent work(\cite{a7} and extensive references therein ), we have studied
the barrier crossing of a time dependent potential which adiabatically evolves in time that sheds some light on the global optimization problem.  

The transport and diffusion properties of Brownian particles continues to attract enormous 
interest and activity even though a century has passed since the appearance of the famous seminal work of Einstein on the subject\cite{b1}. The motion of atoms, vacancies, excitations, molecules, molecular clusters and colloidal particles on surfaces in an active area of research due to its theoretical interest\cite{b2} and modern technological relevance involving self-assembled molecular film growth, catalysis and surface bound nanostructures\cite{1b3}. More recently, the problem of modeling molecular motors\cite{b3},i.e., microscopic objects moving unidirectionally along periodic structures, has renewed the interest in the field and stimulated much theoretical work devoted to the
study of the directed motion in a fluctuating environment in the absence of bias forces.
One of the more recent research foci concerns the experimental and numerical observation that
even for large clusters of molecules, long jumps spanning many lattice sites may in some cases 
be the dominant contributor to the motion\cite{b4}. In recent work, the long, even L$e^{'}$vy-like,
motions can be described by ordinary Langevin dynamics in the low friction regime.
Another related work has dealt with the transport properties of particles in a symmetric periodic potential subject to thermal effects and external forces involving important physical applications
that include Josephson junctions, superionic conductors, colloidal spheres and polymers diffusing at interfaces among many others\cite {b5}.
Several different groups have studied the diffusion of Brownian particle in a periodic field of force\cite{c1}.However, these studies focus on the fact that the motion in periodic potentials is necessarily
diffusive on very long time scales.
Restricting ourselves to the one dimensional case,
the mean-square displacement is given by $<[x(t)-x(0)]^{2}>=2Dt$ for large times $t$. Where $D$ is the diffusion constant.  
One can map this problem onto a quantum system, representing the motion of a quantum particle 
in a modulated or random potential\cite{c9}. On this basis, the long time diffusive properties,
including the discontinuous dependence of the diffusion constant on the 'wavelength' of the inhomogeneities in quasiperiodic media, is related to the leading low frequency behavior 
of the density of states in the associated quantum system. There are the connections between
the classical diffusion, localization, and intermittency\cite{c11}.

In the present work, we study the 
transport of a brownian particle in the field 
of force derivable from a potential with $N$ number of minima with same barrier heights  by using Fokker-Planck dynamics.
We calculate exactly the Kramers' time that the time scale to approach to equilibrium in this case and study how it depends on number of minima. By taking different model potential with $N$ number of minima and of same barrier heights,
we have seen that Kramers time is independent of the shape of the potential. We have seen the localised behavior of the particle in the asymptotic time limit in this case. 
Then we take the limit 
of $N \rightarrow \infty$ to get the periodic potential and study the diffusive behavior.
We calculate exactly the diffusion coefficient in this case in the asymptotic time limit.
We also study the transport in commensurate and incommensurate media.

\section{Localization in the potential with $N$ minima}

Let us consider, a brownian particle is moving in an external potential $V(x)$. The 
overdamped motion of the position variable obeys the Langevin equation
{\fontsize{8pt}{8pt}\selectfont
\bea\label{41ab}
\dot{x}=-\Gamma \frac{\partial V}{\partial x} +\eta
\eea}

Where, $\Gamma$ is the inverse function. $\eta(t)$ is the random force with Gaussian 
distribution and its correlation function obeys 
$<\eta(t_{1})\eta(t_{2})>=2\epsilon \delta(t_{1}-t_{2})$

The time dependent probability distribution $P(x,t)$ of the random force $x(t)$ 
obeys the Fokker Planck equation
{\fontsize{8pt}{8pt}\selectfont
\bea\label{43ab}
\frac{\partial P}{\partial t}=\frac{\partial}{\partial x}(P\frac{\partial V }{\partial x})
+\epsilon \frac{\partial^{2}P}{\partial x^{2}}
\eea}    

which,with substitution,
{\fontsize{8pt}{8pt}\selectfont
\bea\label{44ab}
P(x,t)=\exp\{-\frac{V}{2\epsilon}\} \ \ \phi(x,t)
\eea}

reduces to Schrodinger like equation
{\fontsize{8pt}{8pt}\selectfont
\bea\label{45ab}
-\frac{\partial \phi}{\partial t}=\left[-\epsilon \frac{\partial^{2}}{\partial x^{2}}
+\left(\frac{{V^{\prime}}^{2}}{4\epsilon}-\frac{V^{\prime \prime}}{2}\right)\right]\phi
\eea}

The time independent equation reads,
{\fontsize{8pt}{8pt}\selectfont
\bea\label{46ab}
\left[-\epsilon\frac{\partial^{2}}{\partial x^{2}}+\left(\frac{{V^{\prime}}^{2}}{4\epsilon}
-\frac{V^{\prime \prime}}{2}\right)\right]\psi=\lambda \psi
\eea}

where,$\psi(x,t)=e^{\lambda t}\phi(x,t)$.

\subsubsection{Kramers' time}
\paragraph{Case I:}
Let us start with a model potential with number of minima is two.
Then generalize to $N$ number of minima and periodic potential.

Consider,the potential
\ber\label{48ab}
V(x)&=&V_{0}(1-\frac{x}{a}), \ \ \ \ 0\leq x\leq a \nonumber \\
&=& V_{0}(1+\frac{x}{a}), \ \ \ \ -a \leq x \leq 0 \nonumber \\
&=& V_{0}(\frac{x}{a}-1), \ \ \ \ a\leq x \nonumber \\
&=& -V_{0}(\frac{x}{a}+1), \ \ \ x\leq -a
\eer

Now, the Schrodinger  like  eq.(\ref{46ab}) becomes with the above form of potential
 
\bea\label{49ab}
-\epsilon\frac{d^{2}\psi_{n}}{d x^{2}}+\frac{V_{0}}{a}[\delta(x+a)-\delta(x)+\delta(x-a)]\psi_{n}
=-k^{2}\psi_{n}
\eea
where,
\bea\label{410ab}
\lambda_{n}=\frac{{V_{0}}^{2}}{4a^{2}\epsilon^{2}}-k^{2}
\eea

Consider,
\ber\label{411ab}
\psi_{n}(x)&=&A_{0}e^{-k x}+B_{0}e^{k x},\ \ -a\leq x\leq 0 \nonumber \\
&=&A_{1}e^{-kx}+B_{1}e^{kx},\ \ 0\leq x \leq a \nonumber \\
&=&A_{2}e^{-k(x-a)}, \ \ \ \ \ \ \ \ x\geq a 
\eer
Now, apply the boundary condition at $x=0,a$ namely\\
i)$\psi$ is continuous.\\
ii)$\psi^{\prime}$ is ordinarily continuous i.e. it suffers a discontinuity proportional to 
the strength of the potential.\\
We get the following relation between the coefficients $A_{2}$ at $x \geq a$ and$(A_{0},B_{0})$
at $-a\leq x\leq 0$ .
{\fontsize{8pt}{8pt}\selectfont
\bea\label{412ab}
\left(\begin{array}{c}A_{2}\\0\end{array}\right)=
\left(\begin{array}{c}\begin{array}{cc}e^{-ka}&e^{ka}\\\alpha e^{-ka}&(1+\alpha )e^{ka}
\end{array}\end{array}\right)
\left(\begin{array}{c}\begin{array}{cc}1+\alpha &\alpha\\-\alpha &1-\alpha 
\end{array}\end{array}\right)
\left(\begin{array}{c}A_{0}\\B_{0}\end{array}\right)
\eea}

where,$\alpha=\frac{V_{0}}{2a\epsilon k}$.
$\psi_{n}(x)$ is symmetric arround $x=0$ gives the relation

\bea\label{413ab}
B_{0}=\frac{1+\alpha}{1-\alpha}A_{0}
\eea

With the help of (\ref{412ab}) and (\ref{413ab}) we get the recursion relation of $k$
\bea\label{414ab}
e^{2ka}\frac{V_{0}-2ak\epsilon}{V_{0}}=1
\eea

In order get the first excited state eigenvalue we use
\bea\label{415ab}
k=\frac{V_{0}}{2a\epsilon}+\delta
\eea
and from eq.(\ref{414ab}) we get
\bea\label{416ab}
\delta=-\frac{V_{0}}{2a\epsilon}e^{-V_{0}/\epsilon}
\eea
Now, from (\ref{410ab}), we get the first excited state eigenvalue.
\bea\label{417ab}
\lambda_{1}=\frac{{V_{0}}^{2}}{2a^{2}\epsilon^{2}}e^{-V_{0}/\epsilon}
\eea
In the asymptotic time limit $\lambda_{1}$ gives the dominant contribution in the time dependent 
probability distribution. Hence
the Kramers' time that the time scale to approach to equilibrium is
\bea\label{418ab}
\tau_{0}=\frac{2a^{2}\epsilon^{2}}{{V_{0}}^{2}}e^{V_{0/\epsilon}}
\eea

Now, consider the number of minima is three and 
intervening two maxima are of equal height $V_{0}$.
As in the previous case, Schr\"{o}dinger like equation becomes
\ber\label{419ab}
-\epsilon \frac{d^{2}\psi_{n}}{d\psi^{2}}+\frac{V_{0}}{a}
[\delta(x+2a)-\delta(x+a)+\delta(x)&& \\ \nonumber -\delta(x-a)+\delta(x-2a)]
\psi_{n}=-k^{2}\psi_{n}
\eer
with $\lambda_{n}=\frac{{V_{0}}^{2}}{4a^{2}\epsilon}-k^{2}$.

Consider, 
\ber\label{420ab}
\psi_{n}&=& A_{0}e^{-k x}+B_{0}e^{k x},\ \ \ \ \ \ \ \ \ \ -a\leq x\leq 0 \nonumber \\
&=& A_{1}e^{-k x}+B_{1}e^{k x}, \ \ \ \ \ \ \ \ \ \ 0\leq x\leq a \nonumber \\
&=& A_{2}e^{-k (x-a)}+B_{2}e^{k (x-a)}, \  a\leq x \leq 2a \nonumber \\
&=& A_{3}e^{-k(x-2a)},\ \ \ \ \ \ \ \ \ \ \ \ \ \ \ \ x \geq 2a
\eer

As in the previous case, by applying boundary conditions on $\psi$ and $\psi^{\prime}$
at $x=0,a,2a$, we get the following relation between coefficients $(A_{3},0)$ at $x \geq 2a$ and 
$(A_{0},B_{0})$ at $-a \leq x \leq 0$.
\begin{widetext}
{\fontsize{8pt}{8pt}\selectfont
\bea\label{421ab}
\left(\begin{array}{c}A_{3}\\0\end{array}\right)=
\left(\begin{array}{c}\begin{array}{cc}e^{-ka}&e^{ka}\\\alpha e^{-ka}&(1+\alpha )e^{ka}
\end{array}\end{array}\right)
\left(\begin{array}{c}\begin{array}{cc}
(1+\alpha)e^{-ka}&\alpha e^{ka}\\-\alpha e^{-ka}&(1-\alpha )e^{ka}
\end{array}\end{array}\right)
\left(\begin{array}{c}\begin{array}{cc}1-\alpha &-\alpha\\\alpha &1+\alpha 
\end{array}\end{array}\right)
\left(\begin{array}{c}A_{0}\\B_{0}\end{array}\right)
\eea}
\end{widetext}
From (\ref{421ab}) and (\ref{413ab}),we get the following recursion relation of $k$
{\fontsize{8pt}{8pt}\selectfont
\bea\label{422ab}
\left(\frac{V_{0}}{2a\epsilon k}\right)^{2}e^{2ka}
=-\left(1-\frac{V_{0}}{2a\epsilon k}\right)
\left[1+\left(1-\frac{V_{0}}{2a\epsilon k}-\frac{{V_{0}}^{2}}{2a^{2}\epsilon^{2}k^{2}}\right)
e^{4ak}\right]
\eea}

By substituting $k=\frac{V_{0}}{2a\epsilon}+\delta$ into eq.(\ref{422ab}), we get
\bea\label{423ab}
\delta=-\frac{V_{0}}{4a\epsilon}e^{V_{0}/\epsilon}
\eea
Hence, the Kramers' time in this case is
\ber\label{424ab}
\tau_{1}&=&\frac{4a^{2}\epsilon^{2}}{{V_{0}}^{2}}e^{V_{0}/\epsilon}\nonumber \\
&=& 2\tau_{0}
\eer


Now, we generalize the potential with $N$ minima. The Schr\"{o}dinger like equation
takes the form
\bea\label{425ab}
-\epsilon \frac{\partial^{2}\psi_{n}}{\partial x^{2}}
+\frac{V_{0}}{a}\sum^{N}_{i=-N}(-)^{i+1}\delta(x-(i-1)a)\psi_{n}=-k^{2}\psi_{n}
\eea

The $\psi_{n}$ takes the following form
\ber\label{426ab}
\psi_{n}(x)&=&A_{0}e^{-kx}+B_{0}e^{kx}, \ \ \ \ -a \leq x \leq 0 \nonumber \\ 
 &=& A_{1}e^{-kx}+B_{1}e^{kx}, \ \ \ \ 0\leq x \leq a \nonumber \\
&&\vdots \nonumber \\ 
&=& A_{i}e^{-k(x-(i-1)a)}, \ \ \ \ \ \   x \geq (i-1)a
\eer

As in the previous case, by applying boundary conditions at $x=0,a,\ldots,(i-1)a$, we get 
the following relation between coefficients $(A_{i},0)$ at $x\geq (i-1)a$ and
$(A_{0},B_{0})$ at $-a\leq x\leq 0$.
{\fontsize{8pt}{8pt}\selectfont
\bea\label{427ab}
\left(\begin{array}{c}A_{i}\\0\end{array}\right)
=\left(R\right)\left(T_{1}\right)\left(T_{2}\right)\left(T_{1}\right)
\cdots\left(T_{1}\right)\left(L_{1}\right)\left(\begin{array}{c}A_{0}\\B_{0}\end{array}\right)
\eea}
when $N$ is even. There are $(N-2)$ number of matrices in between $(R)$ and $(L_{1})$ matrices.
and when $N$ is odd, the relation becomes
{\fontsize{8pt}{8pt}\selectfont
\bea\label{428ab}
\left(\begin{array}{c}A_{i}\\0\end{array}\right)
=\left(R\right)\left(T_{1}\right)\left(T_{2}\right)\left(T_{1}\right)
\cdots\left(T_{2}\right)\left(L_{2}\right)\left(\begin{array}{c}A_{0}\\B_{0}\end{array}\right)
\eea}
  
where,${\fontsize{8pt}{8pt}\selectfont(R)=\left(\begin{array}{c}\begin{array}{cc}e^{-ka}&e^{ka}\\\alpha e^{-ka}&(1+\alpha )e^{ka}
\end{array}\end{array}\right)}$, ${\fontsize{8pt}{8pt}\selectfont(T_{1})=\left(\begin{array}{c}\begin{array}{cc}
(1+\alpha)e^{-ka}&\alpha e^{ka}\\-\alpha e^{-ka}&(1-\alpha )e^{ka}
\end{array}\end{array}\right)}$,\\ ${\fontsize{8pt}{8pt}\selectfont(T_{2})=\left(\begin{array}{c}\begin{array}{cc}
(1-\alpha)e^{-ka}&-\alpha e^{ka}\\\alpha e^{-ka}&(1+\alpha )e^{ka}
\end{array}\end{array}\right)}$, ${\fontsize{8pt}{8pt}\selectfont(L_{1})=\left(\begin{array}{c}\begin{array}{cc}1-\alpha &-\alpha\\\alpha &1+\alpha 
\end{array}\end{array}\right)}$, and ${\fontsize{8pt}{8pt}\selectfont(L_{2})=\left(\begin{array}{c}\begin{array}{cc}1+\alpha &\alpha\\-\alpha &1-\alpha 
\end{array}\end{array}\right)}$.

Now, using (\ref{413ab}), we get the recursion relation of $k$ from (\ref{427ab}) and (\ref{428ab})
and then using $k=\frac{V_{0}}{2a\epsilon}+\delta$, we find $\delta$ is of the following form 
\bea\label{429ab}
\delta=-\frac{V_{0}}{2(N-1)\epsilon a}e^{-V_{0}/\epsilon} 
\eea

Hence, the Kramers' time reads
\ber\label{430ab}
\tau_{N}&=&\frac{2(N-1)a^{2}\epsilon^{2}}{{V_{0}}^{2}}e^{V_{0}/\epsilon}\nonumber \\
&=& (N-1)\tau_{0}\nonumber \\
&=& N\tau_{0},\ \ \ \ \ when \ N\gg 1
\eer
\paragraph{Case II:}

Now, we study the approach to equilibrium by using the different model potential.
Consider the potential with minima two is of the form $V(x)=V_{0}(x^{2}-1)^{2}$.
To evaluate the Kramers' time, we have to evaluate the first excited state eigenvalue of the Hamiltonian
of the form $\left[-\epsilon\frac{\partial^{2}}{\partial x^{2}}+\left(\frac{{V^{\prime}}^{2}}{4\epsilon}
-\frac{V^{\prime \prime}}{2}\right)\right]$. It can be evaluated by calculating the ground state
eigenvalue of its super symmetric partner Hamiltonian of the form 
$\left[-\epsilon\frac{\partial^{2}}{\partial x^{2}}+\left(\frac{{V^{\prime}}^{2}}{4\epsilon}
+\frac{V^{\prime \prime}}{2}\right)\right]$\cite{c15,c16}. With the help of trial wave function we can evaluate 
the ground state eigenvalue of the partner Hamiltonian of the order of $e^{-V_{0}}$. Hence the first excited state eigenvalue of the original Hamiltonian is of the order of $e^{-V_{0}}$. 
Now the Kramers' time, the time scale to approach to equilibrium is of the order of $e^{V_{0}}$.

Now, consider the potential with three minima and each maxima is of the same height as double well potential. In the similar way
we can evaluate the Kramers' time of the order of $2{e^{V_{0}}}$, where each maxima gives the same contribution to the Kramers' time.

As a generalization, in case of potential with $'N'$ number of minima and each of the $N-1$ maxima is of the same height as double well potential, the Kramers' time is of the order of $(N-1)e^{V_{0}}$.

Hence the Kramers' time reads
\bea\label{431ab}  
\tau_{N}=N(\tau_{0}),\ \ \ \ \ when \ N\gg 1
\eea
Where, $\tau_{0}\sim e^{V_{0}}$ is the Kramers' time of the potential with minima two.

\subsubsection{Localised to diffusive behavior}
The probability distribution $P(x,t)$ is given by
\bea\label{431ab1}
P(x,t)=P_{eq} +c_{1}\phi_{1}(x)e^{-\lambda_{1}t}+\cdots
\eea
where $\lambda_{1}$ is the first excited state eigenvalue.

The diffusion coefficient is given by
\ber\label{432ab}
D&=&\frac{1}{2}\lim_{t\rightarrow \infty} \frac{\partial}{\partial t}<x^{2}>(t) \nonumber \\
&=& \frac{1}{2}\lim_{t\rightarrow \infty} 
\frac{\partial}{\partial t}\int dx x^{2}P(x,t) \nonumber \\
&=& \frac{1}{2} \lim_{t\rightarrow \infty}
\left[-c_{1}\lambda_{1} e^{-\lambda_{1}t}\int dx x^{2}\phi_{1}(x) +\cdots \right]\nonumber\\
&\rightarrow& 0 
\eer

As long as $N$ is finite, $\lambda_{1} $ is constant. The diffusion coefficient goes to 
zero exponentially in the asymptotic time limit.
Now, as $N \rightarrow \infty$, the potential passes over to periodic potential.
We will see in that case, the transition from localised to diffusive behaviour as 
$t\rightarrow \infty$.
Clearly, as $N\rightarrow \infty$, the excited states eigenvalues tend to zero.
The ground state becomes degenerate and band appears in the periodic potential. 

The probability distribution in this case is given by
\ber\label{432ab1}
P(x,t)&=&P_{eq}(x)+\sum_{n}\int dk \ a_{n} \ e^{ikx} \psi_{n}(x) e^{-D k^{2}t}\nonumber \\
&=& P_{eq}(x)+\sum_{n}a_{n}\psi_{n}(x)e^{-\frac{x^{2}}{4D t}}\sqrt{\frac{\pi}{D t}}
\eer

Where, we have used eigenvalue $\lambda_{n}(k)\sim k^{2}$. $n$ and $k$ are the band and Bloch 
index respectively. Hence the approach to equilibrium distribution $P_{eq}$ is power law type in 
$t$ in case of periodic potential.

\section{Diffusive behavior in the periodic potential}
\subsubsection{Diffusion coefficient}
The Schr\"{o}dinger like equation becomes
\bea\label{433ab}
-\epsilon \frac{d^{2}\psi_{n}}{dx^{2}}+\frac{V_{0}}{a}
\left[\sum^{\infty}_{-\infty}(-)^{n+1}\delta(x-na)\right]\psi_{n}=-k^{2}\psi_{n}
\eea
where, $\lambda_{n}=\frac{{V_{0}}^{2}}{4a^{2}\epsilon}-k^{2}$.

The general solution is 
\ber\label{434ab}
\psi(x)&=& A e^{k(x-2a)}+B e^{-k(x-2a)},\ \ \ \ 2a \leq x\leq 3a \nonumber \\
&=&F e^{k(x-a)}+G e^{-k(x-a)},\ \ \ \ a\leq x \leq 2a 
\eer

According to Bloch's theorem, the wave function in the cell $0\leq x \leq a$ is
\bea\label{435ab}
\psi(x)=e^{-2i k^{\prime}a}\left[A e^{k x}+B e^{-k x}\right]
\eea

At $x=0$, $\psi$ is continuous and $\psi^{\prime}$ suffers the discontinuity proportional 
to the strength of the potential.Hence we get the following relations between coefficients.
\bea\label{436ab}
F=e^{-k a}\left[(1-\frac{V_{0}}{2ak\epsilon})A -\frac{V_{0}}{2ak\epsilon}B\right]
\eea
and
\bea\label{437ab}
G=e^{ka}\left[\frac{V_{0}}{2ak\epsilon}A+(1+\frac{V_{0}}{2ak\epsilon})B\right]
\eea
Similarly,matching at $x=2a$, we get the following relations
\bea\label{438ab}
F=e^{-2ik^{\prime}a}\left[(1-\frac{V_{0}}{2ak\epsilon})Ae^{ka}
-\frac{V_{0}}{2ak\epsilon}Be^{-ka}\right]
\eea
and
\bea\label{439ab}
G=e^{-2ik^{\prime}a}\left[\frac{V_{0}}{2ak\epsilon}Ae^{ka}+
(1+\frac{V_{0}}{2ak\epsilon})Be^{-ka}\right]
\eea

From (\ref{436ab}),(\ref{437ab}),(\ref{438ab}) and (\ref{439ab}) we find the recursion relation of $k$
\bea\label{440ab}
\cos 2k^{\prime}a =p^{2}+(1-p^{2})\cosh 2ka
\eea

Where $k^{\prime}$ is the Bloch index and $k $ is the band index.
$p$ is given by $\frac{V_{0}}{2ak\epsilon}$.
By substituting $k=\frac{V_{0}}{2a\epsilon}+\delta$ in (\ref{440ab}) we find
\bea\label{441ab}
\delta=-\frac{{k^{\prime}}^{2}a V_{0}}{\epsilon}
\frac{1}{2\sinh^{2}V_{0}/2\epsilon}
\eea
From $\lambda_{n}=\frac{{V_{0}}^{2}}{4a^{2}\epsilon^2}-k^2$, we get 
$\lambda_{1}=\frac{{k^{\prime}}^{2}{V_{0}}^{2}}{2\epsilon^{2}}
\frac{1}{\sinh^{2}V_{0}/2\epsilon}$.

Hence, the diffusion coefficient $D$ is given by
\bea\label{442ab}
D=\frac{{V_{0}}^{2}}{2\epsilon^{2}}
\frac{1}{\sinh^{2}V_{0}/2\epsilon}
\eea

Which is independent of period of the potential and only depends on the height of the potential.  It can be written in the following form
\bea\label{443ab}
D=\frac{{V_{0}}^{4}/8\epsilon^{4}}{<e^{V/\epsilon}><e^{-V/\epsilon}>}
\eea

where, $\frac{1}{\sinh^{2}V_{0}/\epsilon}=\frac{{V_{0}}^{2}/4\epsilon^{2}}{<e^{V/\epsilon}><e^{-V/\epsilon}>}$. $<e^{V/\epsilon}>=\frac{1}{2a}\int^{a}_{-a}e^{V/\epsilon}d x$ where, in the 
range $0\leq x\leq a$, $V(x)=V_{0}(1-x/a)$ and in the range $-a\leq x\leq 0$, $V(x)=V_{0}(1+x/a)$.

\subsubsection{Transport in commensurate and incommensurate media }

We consider the potential $V(x)=V_{1}(x)+V_{2}(x)$.
Where, $V_{2}(x+2a)=V_{2}(x)$ and $V_{1}(x+na)=V_{1}(x)$. $V(x)$ has a period $2\pi n$.
We will see how diffusion coefficient depends on $n$.

Consider, the potential $V_{1}(x)$ is of the following form

\ber\label{444ab}
&&V_{1}(x)\nonumber \\&=& V_{0}(1-\frac{x}{a}),\ \ \ \ \  \ \ \ \ 0\leq x \leq a \nonumber \\
&=& -V_{0}(1-\frac{x}{a}),\ \ \ \ \  \ \ \ \ a\leq x \leq 2a \nonumber \\
&=& 3V_{0}(1-\frac{x}{3a}),\ \ \ \ \ \  \ \ \ 2a\leq x\leq 3a \nonumber \\
&=& -3V_{0}(1-\frac{x}{3a}),\ \ \ \ \ \  \ \ \ 3a\leq x\leq 4a \nonumber \\
 &\vdots& \nonumber \\
&=& -(m-2)V_{0}(1-\frac{x}{(m-2)a}), (m-2)a\leq x\leq (m-1)a \nonumber \\
&=& mV_{0}(1-\frac{x}{ma}),\ \ \ \ \ \  \ \ \ (m-1)a\leq x\leq ma \nonumber \\
\eer
and $V_{2}(x)$ is of the following form

\ber\label{445ab}
V_{2}(x)&=& {V_{0}}'(1-\frac{x}{na}),\ \ \ \ \  \ \ \ \ 0\leq x \leq na \nonumber \\
&=&{V_{0}}'(1+\frac{x}{na}),\ \ \ \ \  \ \ \ \ -na\leq x \leq 0 \nonumber \\
\eer

Now, $<e^{V}><e^{-V}>$ takes the following form
\begin{widetext}
{\fontsize{8pt}{8pt}\selectfont
\ber\label{446ab}
&&<e^{V}><e^{-V}>=\nonumber\\ &&\frac{1}{2na}\left\{ \frac{1}{-\frac{V_{0}}{a}-\frac{{V_{0}}'}{na}}
\left[ \frac{1-e^{-2{V_{0}}'}}{1-e^{-2{V_{0}}'/n}}\left ( e^{\frac{2n-2}{n}{V_{0}}'}
-e^{V_{0}+\frac{2n-1}{n}{V_{0}}'}\right)\right] + 
\frac{1}{\frac{V_{0}}{a}-\frac{{V_{0}}'}{na}}
\left[ \frac{1-e^{-2{V_{0}}'}}{1-e^{-2{V_{0}}'/n}}\left (- e^{\frac{2n+2}{n}{V_{0}}'}
+e^{V_{0}+\frac{2n+1}{n}{V_{0}}'}\right)\right]\right\}\nonumber \\&&
\frac{1}{2na}\left\{ \frac{1}{\frac{V_{0}}{a}+\frac{{V_{0}}'}{na}}
\left[ \frac{1-e^{2{V_{0}}'}}{1-e^{2{V_{0}}'/n}}\left ( e^{\frac{-2n+2}{n}{V_{0}}'}
-e^{-V_{0}-\frac{2n-1}{n}{V_{0}}'}\right)\right]  +
\frac{1}{-\frac{V_{0}}{a}+\frac{{V_{0}}'}{na}}
\left[ \frac{1-e^{2{V_{0}}'}}{1-e^{2{V_{0}}'/n}}\left (- e^{-\frac{2n+2}{n}{V_{0}}'}
+e^{-V_{0}-\frac{2n+1}{n}{V_{0}}'}\right)\right] \right\}
\eer}
\end{widetext}


In the large $n$ limit, the potential $V(x)$ still periodic, and we 
find the finite diffusion coefficient.

When $n$ goes to zero, the potential varies very rapidly. Potential becomes 
quasiperiodic and localization of states occur. Particle becomes localised 
in the asymptotic time limit, hence the diffusion coefficient goes to zero.
\section{Conclusion}
In conclusion, we have studied how Kramers' time depends on the number of minima of the model 
potential. We have seen that it is independent of the shape of the potential. The localized behavior in this model potential is studied when the number of minima
is finite. In this case we have seen the diffusion coefficient goes to zero exponentially in the asymptotic time limit.  In the limit of infinite number of minima in our model potential we have seen 
the transition from localized to diffusive behavior. Approach to equilibrium in the case of periodic potential is power law type in time.  In the periodic field of force we have
calculated exactly the diffusion coefficient of the Brownian particle. In case of rational periodic mixing we have seen the diffusive behavior of the Brownian particle. But in the case of irrational mixing we have seen the localized behavior of the Brownian particle.

\section{Acknowledgments}
I wish to express my gratitude to Prof. J. K. Bhattacharjee for his useful comments and  encouragements. Financial support from CSIR is gratefully acknowledged.

\end{document}